\documentclass[14pt,a4paper]{article}

\usepackage{graphicx}
\usepackage[T1]{fontenc}
\usepackage[utf8]{inputenc}
\usepackage{authblk}
\usepackage[]{cite}
\usepackage[margin=1in]{geometry}
\usepackage{hyperref}

\usepackage{sectsty}
\sectionfont{\centering}
\subsectionfont{\centering}
\subsubsectionfont{\centering}
\usepackage{indentfirst}

\newcommand{\projectName}{VisQualdex}
\newcommand{\appName}{VisQual}

\title{\projectName: a comprehensive guide to good data visualization}     

\author[1]{Jan Sawicki}
\author[2]{Michał Burdukiewicz}
\affil[1]{ORCID: 0000-0002-8930-7564, Warsaw University of Technology, jansawickisawicki@gmail.com}
\affil[2]{ORCID: 0000-0001-8926-582X, Institute of Biotechnology and Biomedicine, Autonomous University of Barcelona michalburdukiewicz@gmail.com}

\begin{document}

\maketitle

\begin{abstract}
The rapid influx of low-quality data visualizations is one of the main challenges in today's communication. Misleading, unreadable, or confusing visualizations spread misinformation. 
Furthermore, they fail to deliver their message to the viewer. 
The scale of the problem is big enough that there already exist public fora gathering tens of thousands of users criticizing graphics and charts (reddit.com/r/dataisugly) made with obvious mistakes. 
Current attempts at data visualization appear mostly as simple and overgeneralized checklists, and often lack systematicity and versatility.  
The lack of proper tooling for evaluating data visualization quality further heightens the problem. 

Therefore, this paper proposes VisQualdex, a systematic set of guidelines for static data visualization. 
The codex categorization is based and inspired by the theory of Grammar of Graphics.
It contains dozens of criteria designed to catch various errors and mistakes of different categories and magnitude. 
Furthermore, it has been peer-reviewed and tested by experts of data visualization, data science, graphics design,  information technology and computer science. 

To apply theory in the real world and increase the practical impact of VisQualdex, this contribution also introduces a practical tool. 
The implementation of the guidelines is available in the form of the web server, \url{https://visqual.onrender.com}, developed as a single page application in JavaScript using Vue.js and Material Design principles.
\end{abstract}

\section{Introduction}
The first historical stamps of data visualization reach back to before 17\textsuperscript{th} century.
Its ``golden age'' dates to the second half of 19\textsuperscript{th} century~\cite{friendly2008brief}.
However, it is the 21\textsuperscript{th} that may be called the renaissance of data visualization. 

Human preference to acquire information with visual means~\cite{kaas2014current} in tandem with the time efficiency of conveying information with charts~\cite{6305953} results in massive production of data visualization applied in various fields, including business analysis, Big Data, psychology, journalism, and production process~\cite{weber2012data, sinar2015data}. 
Mass-produced charts, graphs, diagrams, schemes and infographics flood the market and the viewers~\cite{few2007data}. 

The influx of data visualizations is possible thanks to the vast plethora of specialized tools. Although novel software makes visualization easier, they do not ensure the quality of their creations.
Moreover, no modern tools are designed to evaluate the quality of the data visualization. We are still limited to guidelines presented in checklists, lists of questions or suggestions that often fail to fulfill their task
~\cite{catchpoleProblemChecklists2015}
. To fill this void, we propose organized, systematic data visualization guidelines based on state-of-the-art practices, \projectName.

Our methodology allows everyone, from non-specialists to data science experts, to assess the quality of data visualization and pinpoint existing problems. 
Compared to current methods for data visualization evaluation, \projectName\ leads to a more exhaustive and complete evaluation, due to utilizing a systematic, precise and scientifically supported criteria. 
Moreover, the \projectName\ is also available as the companion web application, \appName,  and thus can be easily incorporated in many design workflows.

\subsection{The spectrum of data visualizations}
The focus of \projectName\ is to evaluate the quality of broadly understood static ``data visualizations''. 
The definitions in the literature vary from the ``image that is representative of the raw data''~\cite{azzam2013data} to ``the set of methods for graphically displaying information in a way that is understandable and straightforward''~\cite{tonidandel2015big}. Despite the relative vagueness of these definitions, they capture the essential aim of the data visualization, which is to communicate information in a graphical form. 
The ``static'' keyword indicated that this guideline refers to visualization that could be simply printed out without losing its key features, i.e. interactive dashboard or real 3D visualization are out of the scope.





\begin{figure}[!htbp]
    \centering
    \includegraphics[width=\textwidth]{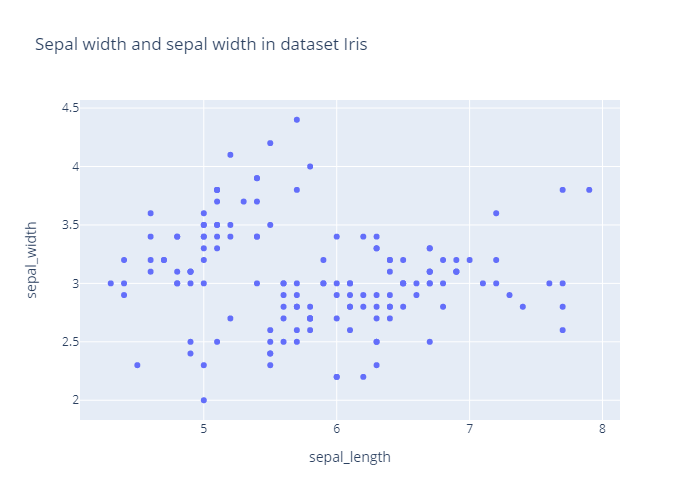}
    \caption{An example of a ``classical'' data visualization}
    \label{fig:simple_chart}
\end{figure}

\begin{figure}[!htbp]
    \centering
    \includegraphics[width=0.8\textwidth]{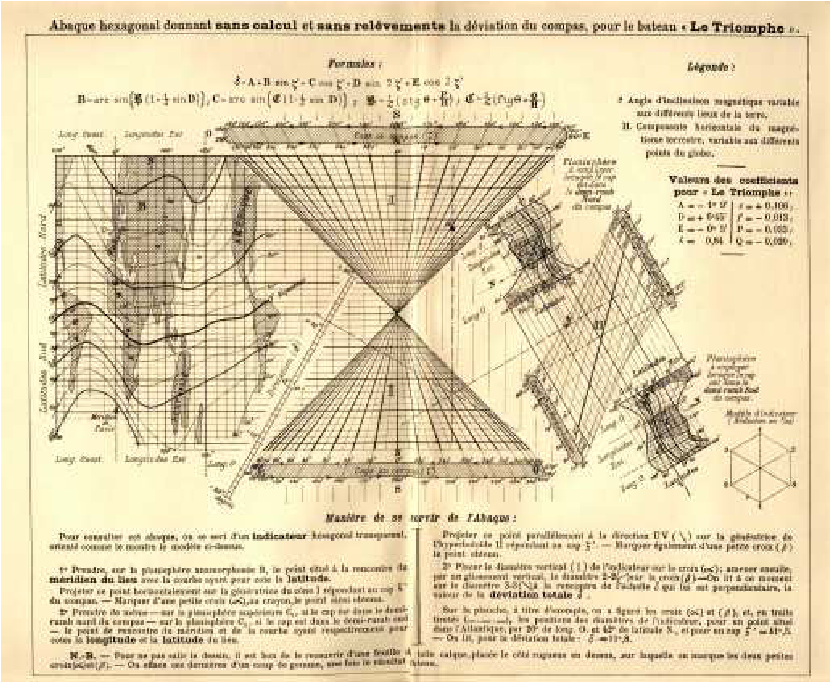}
    \caption{An example of a ``non-classical'' data visualization (adapted from Charles Lallemand's ``tour de force'' nomogram from a work on hexagonal charts~\cite{lallemand1885abaques}}
    \label{fig:alternative_data_visualization}
\end{figure}

In all types of communication, there are many possibilities of conveying the same message. Therefore, the data visualization itself covers drastically different entities, ranging from the simple chart (Figure~\ref{fig:simple_chart}) to more complex visualizations (Figure~\ref{fig:alternative_data_visualization}). It provides an additional layer of complication to the data visualization assessment, as the rules must be general enough to apply to all types of data visualization.


\subsection{State of the art}

Despite the vagueness of definitions, the way we describe data visualizations is subject to numerous improvements over time
~\cite{friendly2008brief}
. These small steps result in more structured descriptions of visualizations, exemplified by the grammar of graphics. These theoretical achievements have contributed to the development of practical tools such as Matplotlib, Seaborn, Plotly and others~\cite{porcu2018matplotlib,waskom2021seaborn,wickham2006introduction,ali2016big}.

In a glaring difference, the evaluation of data visualization is still as unsystematic as it used to be in the past. The first debate on that topic dated back to 1857 and resulted in several advises~\cite{palsky1999debate}. The theoretical understanding of the correctness of data visualization has grown over time with new or revised good practices~\cite{azzam2013data}
, but very few propose practical and applicable evaluation methods.
 
The lack of a systematic approach stems from the popularity of checklists as one of the most commonly adapted evaluation systems. Here, a checklist is a list of potential mistakes, sometimes divided in thematic sections. The first checklists (or guidelines) date back to 1915~\cite{first-data-visualization-guideline}, but the community of data visualization practitioners is still producing newer counterparts
~\cite{evergreen2016data}
. 


Checklists have two main practical advantages: simplicity and shortness. Simplicity means that the majority of the state-of-the-art checklists can be easily operated by any viewer with a basic understanding of the most trivial data visualization concepts, like chart, axis or scale. Moreover, most visualization checklists are brief (e.g., about 20 questions~\cite{evergreen2016data,berkley-checklist}
). This short length, along with the simplicity, translates to a relatively good assessment pace. 

\subsection{Disadvantages of the checklist usage}








However, checklists are characterized by the disorder happening on two different conceptual levels: the lack of organization and varying levels of universality. 

In the majority of studied examples, checklists do not possess any grouping or hierarchy of the guidelines. Although some checklists demonstrate some degree of guideline categorization, they often do not reflect the state-of-art data visualization descriptions and may have a practical rationale. 

The varying level of universality happens when general guidelines (e.g., keep the graph two-dimensional) occur along with more specific pieces of advice (e.g., 'use bar charts to visualize achievement of an objective'). It limits the scope of the checklist to a particular set of data visualizations. This problem is even more pronounced if the checklist contains a scale based on the number of questions answered correctly or fulfilled guidelines. In this situation, non-general guidelines falsely lower the actual score of a visualization.

Moreover, usage of checklists forces following the state-of-the-art methodology and logic of checklists~\cite{scriven2000logic}, which state that (some points are skipped):
\begin{itemize}
    \item ``The list should be complete (no significant omissions).'' and ``The checkpoints should refer to criteria and not mere indicators.''
    
    According to this research, the second rule is most often broken as none of the state-of-the-art checklists have ``complete''/``full'' coverage of the evaluation criteria. 
    As most of both state-of-the-art data visualization checklists aim at an ``engineering'' approach, they tend to have very strict, concise and precise points. 
    However, this causes them to sometimes be superficial, oversimplified and focus on indicators instead of criteria. 
    For example, a rule ``No more than 3 colors''~\cite{berkley-checklist} is focusing purely on indicators, but disallows great 4-color visualizations from passing the benchmark. 
    Additionally, a rule ``Did you start the Y-axis at 0?'' (answer yes/no)~\cite{david-mckie-checklist} allows only charts which do not cut the axis in a justifiable way (e.g. shoe sizes for adults, Earth temperatures with Kelvin units, etc.). 
    \item ``The criteria should be commensurable.''
    
    This criterion is not applicable to data visualization checklists, because some mistakes are more significant than others. 
    There exist visualizations which can follow all criteria except one and still be condemning wrong. 
    \item ``The list should be concise (to assist its mnemonic function).''
    
    In contrast to engineering processes or medical applications
    ~\cite{grigg2015smarter}
    , data visualization does not always follow strict regularities as it is a mix of applied arts and fine arts which cannot be separated
    ~\cite{mirahan7chapter, cairo2012functional}
    . 
\end{itemize}

Therefore, applying checklist format to data visualization evaluation may lead to problems which are neither the problem of the checklist methodology, neither the data visualization evaluation. 
The problem lies in forcefully fusing the two ideas together. 

All problems described above, together with the brevity, result in the non-exhaustiveness of checklists. Right now, there are no checklists that would approach the evaluation of data visualization in a systematized way. Therefore, we have designed \projectName\ to at least partially alleviate these issues and produce an evaluation methodology applicable to the broad spectrum of data visualizations. 


\section{Design of \projectName}
\label{sec:design}
The development of \projectName\ follows the VISupply framework for design of data visualization guidelines~\cite{engelke2018visupply}. It covers four main steps:

\begin{itemize}
    \item Evidence collection \\ 
    Collection and diligent analysis of research works about data visualization. 
    \item Integration \\
    Curation of existing data visualization guidelines, good practices, suggestions and similar.
    \item Contextualization \& Generalization \\
    Merging concepts from different works and forming clusters, a.k.a. ``categories'' (see section \ref{sec:categories}).
    \item Guideline Definition \\
    Formalization of \projectName, i.e. stating the ``question'' format, supplementing missing areas and verifying \projectName\ in practice, utilizing and extending nomenclatural notions
    ~\cite{diehl2020studying}
    .
\end{itemize}


Moreover, an original concept introduced in \projectName\ consists of the four main traits of a correct data visualization.
These four pillars of \projectName\ are:
\begin{enumerate}
    \item Real data instead of guesstimates.
    \item Clarity and readability instead of incomprehensibility and ambiguity.
    \item Simplicity and summarization instead of complexity and raw data.
    \item Guidance and objectivity instead of manipulation and subjectivity.
\end{enumerate}

Finally, the codex has been peer-reviewed by 4 independent reviewers. 
The reviewers were experts and specialists in the following fields (parenthesis contain the reviewer's higher domain): data visualization (computer science), data science (computer science), graphics and design (fine arts) and information technology (computer science).
The reviewers all submitted their critique to all the questions and overall codex design. 
The feedback was gathered in 1–3 iterations, depending on the reviewer. 
All the comments and suggestions caused various criteria (\projectName\ questions) to be introduced, redefined or abandoned due to lack of quality evidence. 

It is important to note that in some initial stages of development, the tool was supposed to be based on user testing and feedback in a ``wisdom of the crowds'' methodology. 
However, the user's feedback was very fragmented and not unanimous. 
Moreover, it was highly biased to the user experience. 
Therefore, the expert approach was chosen to strengthen the final criteria, maximize the good practices and minimize the bias of random user evaluation. 


\subsection{Categories}
\label{sec:categories}
\label{categories}

The important part of \projectName\ is the categorization of guidelines. The baselines are the formalized descriptions of Grammar of Graphics (GoG)~\cite{wilkinson2012grammar}
and Layered Grammar of Graphics (LGoG)~\cite{wickham2010layered}. Although both of these approaches constitute an in-depth description of data visualization, they are used primarily for either building or decomposing the visualization object. As this is a different goal from the evaluation of data visualizations, GoG and LGoG are only reference points. Therefore, \projectName\ utilizes a redesigned categorization of guidelines. Furthermore, proposed categories are complete (i.e. there is no ``others'' category) and disjoint (i.e. each of the questions belongs exactly to one category).

\textit{Subjective} \\
The objective of this category is to incorporate any purely subjective aspects of the visualization. Although all categories concern issues that may be answered differently depending on the viewer, this one focuses on things exclusively related to the opinion of the on-looker. 

\textit{Theme} \\
This category contains all visual features and artistic choices not directly depending on data, like colors (not related to the color scale), fonts, spacing, and any additional graphics that are not strictly part of the chart. 

\textit{Coordinates} \\
This category is responsible for the coordinate system and units. Its purpose is to check if all coordinates systems (or their alternatives), units and axes are correctly prepared, provided and presented. It also examines if the relation between shear data and all the aspects above is consistent.  



\textit{Geometry} \\
This category includes all information about the shapes used for data presentations (e.g., the shape itself and its dimensions). 
It concerns the shape of the whole figure as well as all used figures and any other geometrical aspects. 

\textit{Guides} \\
This category handles any text content that appears on the visualization. It focuses only on the content, not the display of e.g., title, legend, axes labels, additional comments, labels etc.
It verifies the most importantly the content of the text but also its clarity, objectivity and overall necessity. 


\textit{Perception} \\
This category focuses on the general perception of the data. It is also responsible for detecting all misuses leading to the incorrect understanding of the data, e.g., bar charts with bars starting at an arbitrary point to make the difference between bar length more pronounced.

\textit{Data} \\
This category is responsible for evaluating issues related only to data and all the possible issues such as data source/validity, missing data, and appropriateness of data explanations (e.g., used metrics).

\subsubsection{Questions}
\label{sec:questions}

Each category contains questions which represent unitary criteria based on the pillars described in section~\ref{sec:design}. 
The most important features of them are:
\begin{itemize}
    \item All questions are ``yes or no'' and trigger (negative answer) only if something is incorrect. 
    \item Questions do not overlap or include each other. 
    \item All questions address as general issues as possible while focusing on one particular type of mistake. 
    It means that each can question can be applied to any visualization regardless of factors such as form, type, content. 
    However, some categories are incompatible with some visualizations by definition, e.g., a simple bar chart without any faceting cannot be evaluated in terms of faceting. 
    \item It is possible for one general bad practice to trigger many questions
    \item Depending on the context, a single negative answer may have a tiny or gigantic impact on the visualization understanding.
    It means that it is impossible to judge the quality of a visualization solely by the fraction of positively answered questions.
    \item Literature sources support most questions (the complete citations list available in the supplementary materials). 
\end{itemize}

\begin{figure}[!htbp]
    \centering
    \includegraphics[width=0.6\textwidth]{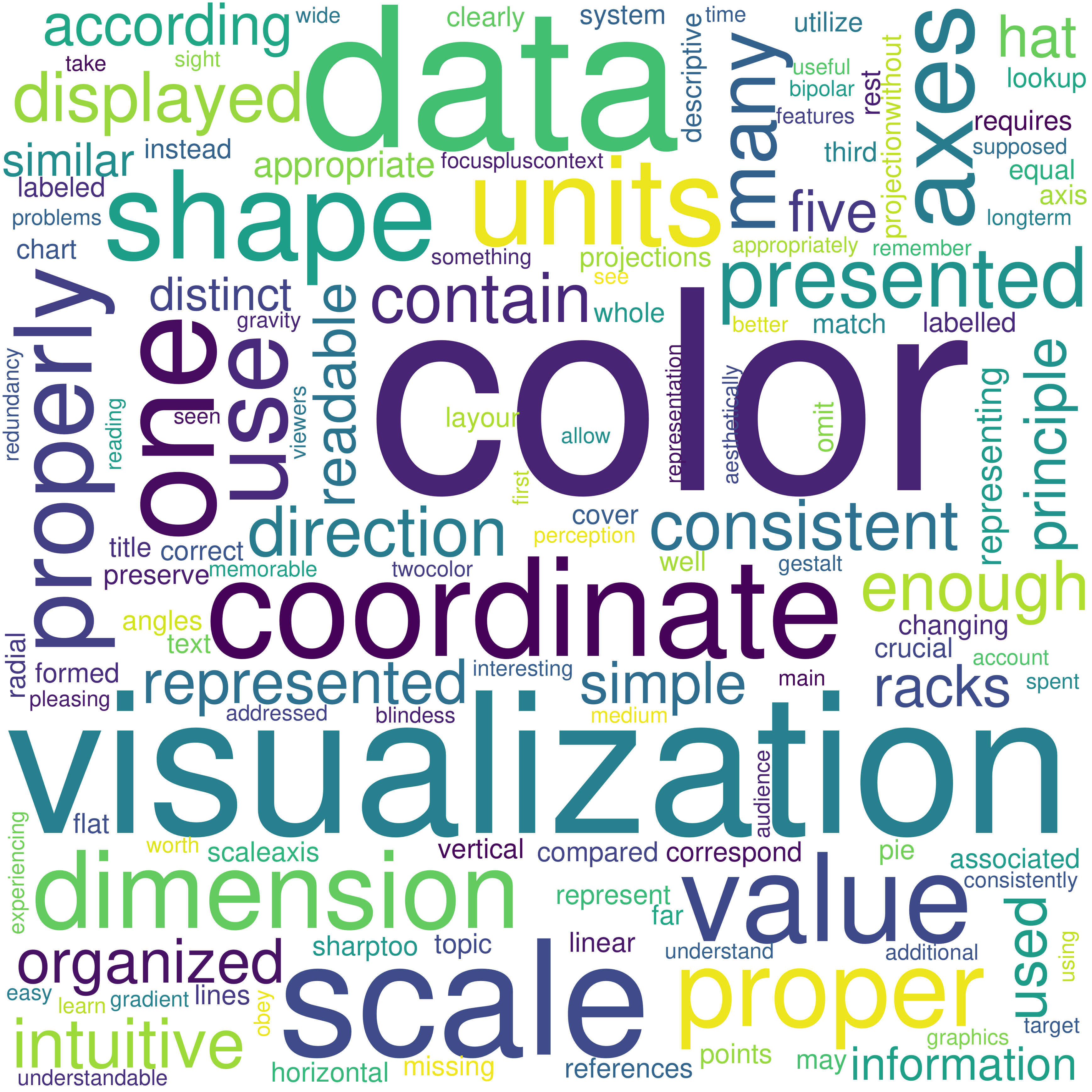}
    \caption{Wordcloud displaying most popular words used in questions (without stopwords)}
    \label{fig:question_wordcloud}
\end{figure}

\projectName\ contains a total of 60 criteria in the form of a question which address/detect different mistakes. We present their general content in the form of the word cloud (Figure~\ref{fig:question_wordcloud}). 

The full list of questions is available in the supplementary materials. 
Here, we present and analyze a few exemplary questions.

\begin{figure}[!htbp]
    \centering
    \includegraphics[width=\textwidth]{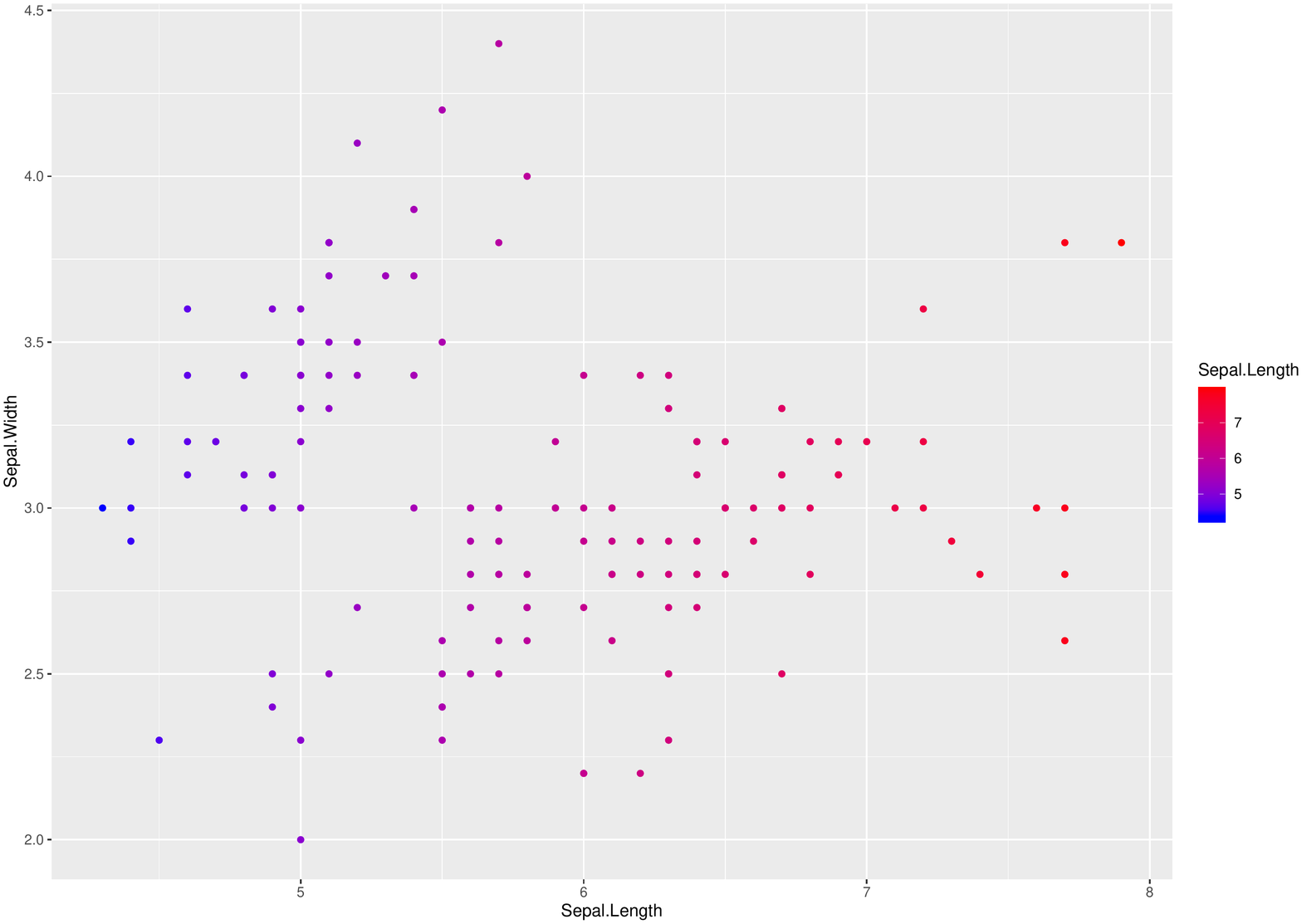}
    \caption{Example of an equidistributed color gradient scale (using dataset Iris~\cite{fisher1936use})}
    \label{fig:ColorGradientScale}
\end{figure}

\textbf{Q}: "Is there not too many colors representing the data?"
In the case of gradient color scale, the distribution of the colors should be regular. 
The figure~\ref{fig:ColorGradientScale} shows a proper distribution of colors on a gradient scale. 
Even though this rule does not touch upon the topic of the choice of colors, it is worth mentioning that other studies~\cite{rogowitz1999trajectories}
 suggest refraining from ``rainbow scale'' and advise simpler/fewer color combinations instead. 

\begin{figure}
    \centering
    \includegraphics[width=1.0\textwidth]{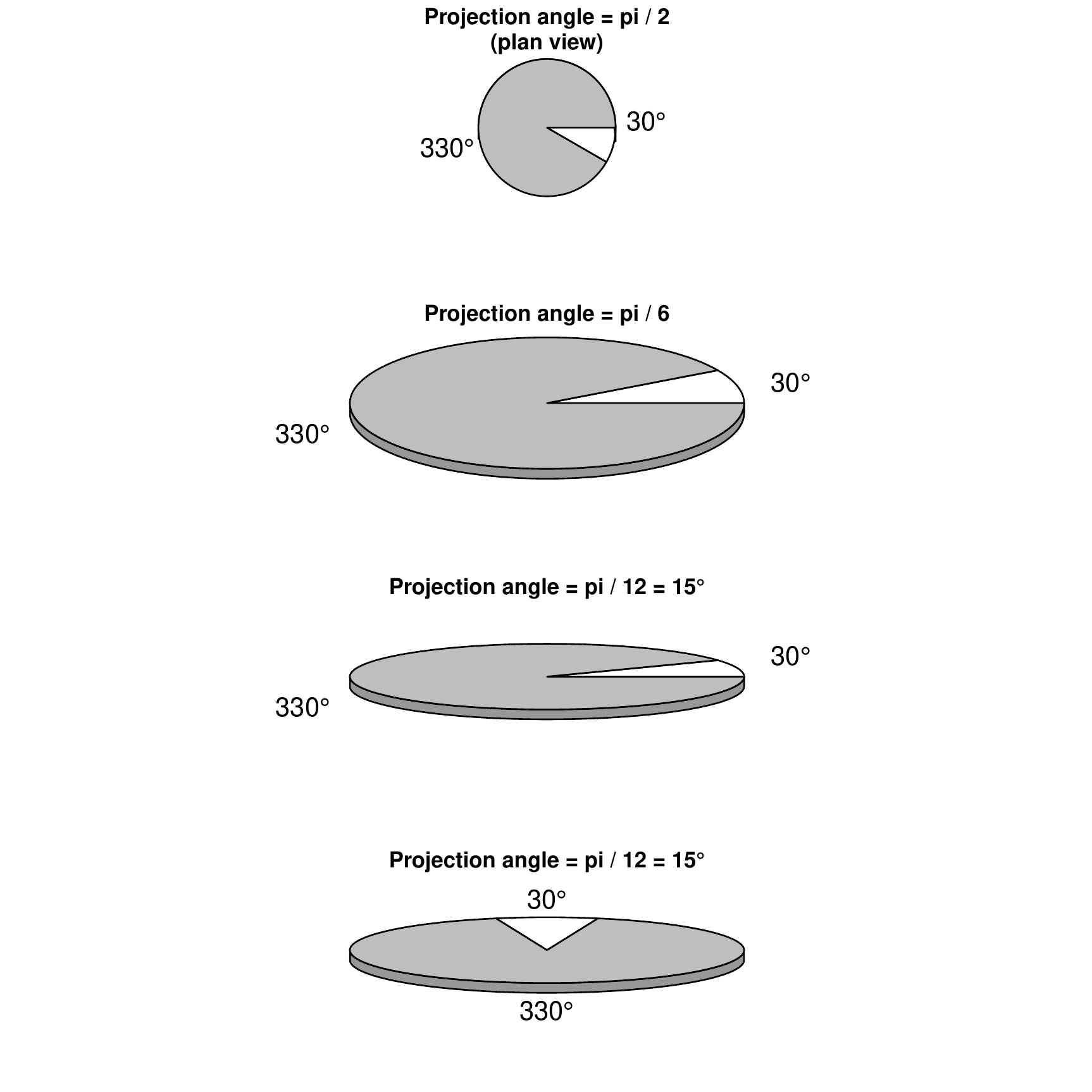}
    \caption{Example of different angles for a 3D pie chart}
    \label{fig:3dPieChartsAngles}
\end{figure}

\textbf{Q}: "Does it omit or utilize properly the third dimension?",
This question focuses on minimizing the additional dimensional complications of the visualization. 
According to state-of-the-art research~\cite{bertini2016judgment} using more than two dimensions on a visualization may be misleading and difficult to perceive. 
Moreover, another study~\cite{cleveland1985graphical} shows that interpreting angles (which appear a lot more often on 3D graphics) has many possible vicious implications, ranging from minor ``illusory effects'' distorting the viewer perception to completely hiding some data points on the visualization. 
See figure~\ref{fig:3dPieChartsAngles} for a graphical example of how differently the same angle (data point) looks due to different projections.

\textbf{Q}: "Does the visualization obey the reading gravity?",
The concept of reading gravity~\cite{reading-gravity} comprehensively aggregates all aspects of the order in which the viewer perceives (reads) the visualization. 
It accounts for how the user reads the text of the visualization and in what order they see the data points, sub-charts (if faceting) and all other visuals. 
Although most western cultures are sinistrodextral (i.e. read from left to right) and from top to bottom, some cultures follow different patterns (e.g., Arabic is written right to left
~\cite{jordan2014reading}
, Hanunuo script is written bottom to top
~\cite{epo2014discourse}
). 
Hence, to maintain flexibility and universality, the question imposes general ``reading gravity'' instead of ``left to right, top to bottom''. 

\section{Technical implementation}

\begin{figure}[!htbp]
    \centering
    \includegraphics[width=\textwidth]{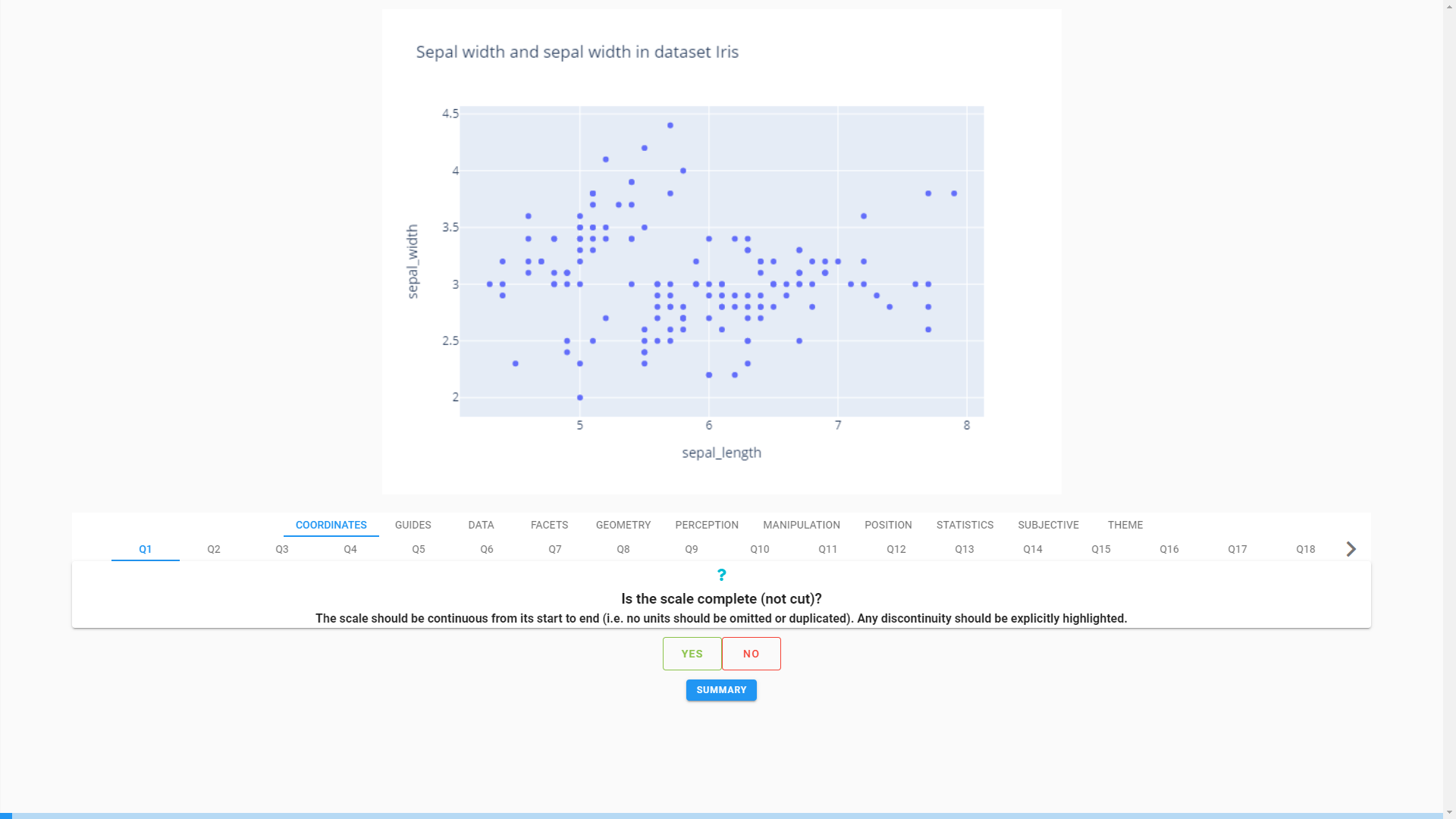}
    \caption{VisQual website screenshot}
    \label{fig:visqual-screenshot}
\end{figure}

The \projectName has been implemented as an easy to use online tool for data evaluation.
The tool is hosted at \url{https://visqual.onrender.com} and a screenshot is presented in Figure \ref{fig:visqual-screenshot}. 
The tool allows anyone, even without advanced visualization knowledge, to upload a data evaluation (a picture in any common format, i.e. JPG, PNG etc.) and evaluate it by going through all the questions. 
Each question has a yes or no answer. 
If a question does not apply to a particular chart or a user cannot answer a question because they are unsure, they can skip it, which marks them as ``NA'' (Not Applicable). 
Moreover, most of the questions have examples presenting good or bad practices of that question. 
For example, the question ``Are colors distinct?'' shows a ``bad example'' in which the colors are used incorrectly (i.e. continuous color scale is applied to the categorical axis).
After processing all the questions, a user can see the report with percent of questions answered positively in each category. 
The higher the percent, the better the evaluation is. 
Moreover, they can download a persistent form of this report to the computer in a JSON or CSV format. 

From the technical point of view, the website is implemented in Vue.js~\cite{vue-js} as a single page application (SPA).
Therefore, it does not need to reload the page at any point. 
However, it is responsive to changes and intuitive because it is based on the Material Design~\cite{noauthororeditor2015material}. 
This makes it easy to use for users without technical knowledge. 
Additionally, all questions are stored in an external file, so it is possible to hot swap any content.  
Finally, the whole project has been carefully maintained using a version control system, Git~\cite{6188603}.

\section{Discussion}

The field of data visualization does not suffer from the lack of guidelines and checklists, but rather from their surplus and disorder. This disarray results in a situation where our checklists are not comprehensive and thus dismiss the impact of the interaction of many factors determining the effectiveness of visual communication
~\cite{kandogan2016grounded}. 
\projectName\ is the first codex (a structured set of criteria) that could be at the same time versatile and extensive enough to cover all existing data visualizations. 

Our systematic approach results in a set of rules that constitutes a foundation for tools for data visualization creation (e.g., Microsoft Excel, Plotly, ggplot, Matplotlib, D3, etc.) and instruments for automatic/semi-automatic data visualization correction (e.g., project ReVision~\cite{revision}). A principal example is the usage \projectName\ for default settings of these tools.

One of the problems concerning checklists is a varying level of detail. 
Specific points may mention at the same time significantly narrow and very general criteria while being on the same ``level'' of evaluation or even share the evaluation weight. It leads to imbalanced evaluation, which may either allow ``incorrect'' visualizations to slip through the metric or ``good'' visualizations to be unfairly punished for minor mistakes. 
\projectName\ partially solves this problem with categories, which guard question overlap and thematic division. However, we see it only as one of the first steps into a comprehensive visualization ontology, focused on the evaluation, instead of creation~\cite{duke2004building}
. 

There are certain aspects of data visualization that could not be included in questions due to lack of scientific consensus and ongoing heated debate regarding the right answer.
A primary example of it is the question ``Is the data-to-ink ratio rational?''. 
There are respected experts like Edward Tufte
~\cite{tufte1990envisioning}
and others~\cite{mcgurgan2015data, inbar2007minimalism} who favor minimalism in data visualization and reject ``chart junk''~\cite{gatto2015making}). 
There are also respected experts like Alberto Cairo~\cite{cairo2012functional} who claim that ``chart junk'' can be useful~\cite{li2014chart} and claim that redundancy (e.g., highlighting in color) may help to quicker convey the message and improve memorability~\cite{bateman2010useful}. 
Overly encumbering the visualization with unnecessary information may lead to confusion, but leaving as little trace of the information may also turn a visualization into a ``clue hunt'' instead of quickening information perception. 
A similar debate considers the ``Y axis trimming''~\cite{correll2020truncating}. 
There are works claiming the starting the Y axis from 0 is the best way~\cite{king2018preparing}, while others suggest that different ranges apply in different situations~\cite{witt2019graph}. 
These and other scientific arguments prove that data visualization is first a still developing and lively domain and second, that it is not purely an applied art/exact science but also fine arts/humanities. Moreover, it also implies that the consensus regarding specific aspects of visualizing information is still fluid and in future there will be a need to update the VisQualdex guidelines.

Finally, the current and future techniques of image processing will make it possible to automate or semi-automate some evaluation steps. 

\section{Conclusion}
The following conclusions emerged from this thorough analysis and after seeing thousands of graphs, charts or infographics. 
Data visualization is a field on the border of computer science, data science and arts
~\cite{mirahan7chapter}
, which renders it highly subjective to the bias of the creator and the viewer. 
However, we believe it is possible to forge universal criteria and find a universal standard to visualize data more understandably. \projectName\ is our first attempt at this task.

\appendix
\section{Supplement}
Supplementary materials are in file Supplement 1 - VisQual questions.

\bibliography{bib}{}

\begin{thebibliography}{10}

\bibitem{friendly2008brief}
M.~Friendly, ``A brief history of data visualization,'' in {\em Handbook of
  data visualization}, pp.~15--56, Springer, 2008.

\bibitem{kaas2014current}
J.~H. Kaas and P.~Balaram, ``Current research on the organization and function
  of the visual system in primates,'' {\em // Eye and brain}, vol.~6, no.~Suppl
  1, p.~1, 2014.

\bibitem{6305953}
G.~V. R. J.~S. {Prasad} and A.~{Ojha}, ``Text, table and graph -- which is
  faster and more accurate to understand?,'' in {\em 2012 IEEE Fourth
  International Conference on Technology for Education}, pp.~126--131, 2012.

\bibitem{weber2012data}
W.~Weber and H.~Rall, ``Data visualization in online journalism and its
  implications for the production process,'' in {\em 2012 16th International
  Conference on Information Visualisation}, pp.~349--356, IEEE, 2012.

\bibitem{sinar2015data}
E.~F. Sinar, ``Data visualization,'' {\em // Big data at work: The data science
  revolution and organizational psychology}, pp.~115--157, 2015.

\bibitem{few2007data}
S.~Few and P.~Edge, ``Data visualization: past, present, and future,'' {\em //
  IBM Cognos Innovation Center}, 2007.

\bibitem{catchpoleProblemChecklists2015}
K.~Catchpole and S.~Russ, ``The problem with checklists,'' {\em // BMJ Quality
  \& Safety}, vol.~24, pp.~545--549, Sept. 2015.

\bibitem{azzam2013data}
T.~Azzam, S.~Evergreen, A.~A. Germuth, and S.~J. Kistler, ``Data visualization
  and evaluation,'' {\em // New Directions for Evaluation}, vol.~2013, no.~139,
  pp.~7--32, 2013.

\bibitem{tonidandel2015big}
S.~Tonidandel, E.~B. King, and J.~M. Cortina, {\em Big data at work: The data
  science revolution and organizational psychology}.
\newblock Routledge, 2015.

\bibitem{lallemand1885abaques}
C.~Lallemand, ``Les abaques h{\'e}xagonaux: Nouvelle m{\'e}thode
  g{\'e}n{\'e}rale de calcul graphique, avec de nombreux exemples
  d'application,'' {\em // Minist{\`e}re des travaux publics, Comit{\'e} du
  nivellement g{\'e}n{\'e}ral de la France, Paris}, 1885.

\bibitem{porcu2018matplotlib}
V.~Porcu, ``Matplotlib,'' in {\em Python for Data Mining Quick Syntax
  Reference}, pp.~201--234, Springer, 2018.

\bibitem{waskom2021seaborn}
M.~L. Waskom, ``Seaborn: statistical data visualization,'' {\em // Journal of
  Open Source Software}, vol.~6, no.~60, p.~3021, 2021.

\bibitem{wickham2006introduction}
H.~Wickham, ``An introduction to ggplot: An implementation of the grammar of
  graphics in r,'' {\em // Statistics}, 2006.

\bibitem{ali2016big}
S.~M. Ali, N.~Gupta, G.~K. Nayak, and R.~K. Lenka, ``Big data visualization:
  Tools and challenges,'' in {\em 2016 2nd International Conference on
  Contemporary Computing and Informatics (IC3I)}, pp.~656--660, IEEE, 2016.

\bibitem{palsky1999debate}
G.~Palsky, ``The debate on the standardization of statistical maps and diagrams
  (1857-1901). elements of the history of graphical semiotics,'' {\em //
  Cybergeo: European Journal of Geography}, 1999.

\bibitem{first-data-visualization-guideline}
``Joint committee on standards for graphic presentation,'' {\em // Publications
  of the American Statistical Association}, vol.~14, no.~112, pp.~790--797,
  1915.

\bibitem{evergreen2016data}
S.~Evergreen and A.~K. Emery, ``Data visualization checklist,'' 2016.

\bibitem{berkley-checklist}
B.~D.~I. Team, ``Data visualization checklist,'' {\em // Berkley Business
  Process Management}, pp.~1--4, 2020.

\bibitem{scriven2000logic}
M.~Scriven, ``The logic and methodology of checklists,'' 2000.

\bibitem{david-mckie-checklist}
D.~McKie, ``Data visualization checklist.''
  (http://www.davidmckie.com/infogram-data-visualization-checklist.pdf),
  accessed 2020.

\bibitem{grigg2015smarter}
E.~Grigg, ``Smarter clinical checklists: how to minimize checklist fatigue and
  maximize clinician performance,'' {\em // Anesthesia \& Analgesia}, vol.~121,
  no.~2, pp.~570--573, 2015.

\bibitem{mirahan7chapter}
M.~Mirahan-Farag, ``Chapter fourteen the segregation of applied arts from fine
  arts and the status of fashion,'' {\em // An Anthology of Philosophical
  Studies Volume 7}, p.~145.

\bibitem{cairo2012functional}
A.~Cairo, {\em The Functional Art: An introduction to information graphics and
  visualization}.
\newblock New Riders, 2012.

\bibitem{engelke2018visupply}
U.~Engelke, A.~Abdul-Rahman, and M.~Chen, ``Visupply: A supply-chain process
  model for visualization guidelines,'' in {\em 2018 International Symposium on
  Big Data Visual and Immersive Analytics (BDVA)}, pp.~1--9, IEEE, 2018.

\bibitem{diehl2020studying}
A.~Diehl, M.~Kraus, A.~Abdul-Raman, M.~El-Assady, B.~Bach, R.~S. Laramee,
  D.~Keim, and M.~Chen, ``Studying visualization guidelines according to
  grounded theory,'' {\em // arXiv preprint arXiv:2010.09040}, 2020.

\bibitem{wilkinson2012grammar}
L.~Wilkinson, ``The grammar of graphics,'' in {\em Handbook of computational
  statistics}, pp.~375--414, Springer, 2012.

\bibitem{wickham2010layered}
H.~Wickham, ``A layered grammar of graphics,'' {\em // Journal of Computational
  and Graphical Statistics}, vol.~19, no.~1, pp.~3--28, 2010.

\bibitem{rogowitz1999trajectories}
B.~E. Rogowitz, A.~D. Kalvin, A.~Pelah, and A.~Cohen, ``Which trajectories
  through which perceptually uniform color spaces produce appropriate colors
  scales for interval data?,'' in {\em Color and Imaging Conference},
  vol.~1999, pp.~321--326, Society for Imaging Science and Technology, 1999.

\bibitem{fisher1936use}
R.~A. Fisher, ``The use of multiple measurements in taxonomic problems,'' {\em
  // Annals of eugenics}, vol.~7, no.~2, pp.~179--188, 1936.

\bibitem{bertini2016judgment}
E.~Bertini, N.~Elmqvist, and T.~Wischgoll, ``Judgment error in pie chart
  variations,'' in {\em Proceedings of the Eurographics/IEEE VGTC conference on
  visualization: Short papers}, pp.~91--95, 2016.

\bibitem{cleveland1985graphical}
W.~S. Cleveland and R.~McGill, ``Graphical perception and graphical methods for
  analyzing scientific data,'' {\em // Science}, vol.~229, no.~4716,
  pp.~828--833, 1985.

\bibitem{reading-gravity}
C.~Wheildon, D.~Ogilvy, and G.~Heard, {\em Type \& Layout: Are You
  Communicating Or Just Making Pretty Shapes}.
\newblock Kickstarting Business Series, Worsley Press, 2005.

\bibitem{jordan2014reading}
T.~R. Jordan, A.~A. Almabruk, E.~A. Gadalla, V.~A. McGowan, S.~J. White,
  L.~Abedipour, and K.~B. Paterson, ``Reading direction and the central
  perceptual span: Evidence from arabic and english,'' {\em // Psychonomic
  bulletin \& review}, vol.~21, no.~2, pp.~505--511, 2014.

\bibitem{epo2014discourse}
Y.~J.~S. Epo, {\em Discourse analysis of suyot: a Hanunuo-Mangyan folk
  narrative}.
\newblock PhD thesis, Citeseer, 2014.

\bibitem{vue-js}
V.~developers, ``Vue.js -- the progressive javascript framework v3.0..''
  (\url{https://vuejs.org/guide/introduction.html}), 2014.

\bibitem{noauthororeditor2015material}
{Google Inc.}, ``Material design lite,'' 2015.

\bibitem{6188603}
D.~Spinellis, ``Git,'' {\em // IEEE Software}, vol.~29, no.~3, pp.~100--101,
  2012.

\bibitem{kandogan2016grounded}
E.~Kandogan and H.~Lee, ``A grounded theory study on the language of data
  visualization principles and guidelines,'' {\em // Electronic Imaging},
  vol.~2016, no.~16, pp.~1--9, 2016.

\bibitem{revision}
M.~Savva, N.~Kong, A.~Chhajta, L.~Fei-Fei, M.~Agrawala, and J.~Heer,
  ``Revision: Automated classification, analysis and redesign of chart
  images,'' in {\em Proceedings of the 24th Annual ACM Symposium on User
  Interface Software and Technology}, UIST '11, (New York, NY, USA),
  p.~393–402, Association for Computing Machinery, 2011.

\bibitem{duke2004building}
D.~J. Duke, K.~W. Brodlie, and D.~A. Duce, ``Building an ontology of
  visualization,'' in {\em IEEE Visualization 2004}, pp.~7p--7p, IEEE, 2004.

\bibitem{tufte1990envisioning}
E.~R. Tufte, N.~H. Goeler, and R.~Benson, {\em Envisioning information},
  vol.~2.
\newblock Graphics press Cheshire, CT, 1990.

\bibitem{mcgurgan2015data}
K.~McGurgan, {\em Data-ink ratio and task complexity in graph comprehension}.
\newblock Rochester Institute of Technology, 2015.

\bibitem{inbar2007minimalism}
O.~Inbar, N.~Tractinsky, and J.~Meyer, ``Minimalism in information
  visualization: attitudes towards maximizing the data-ink ratio,'' in {\em
  Proceedings of the 14th European conference on Cognitive ergonomics: invent!
  explore!}, pp.~185--188, 2007.

\bibitem{gatto2015making}
M.~A. Gatto, ``Making research useful: Current challenges and good practices in
  data visualisation,'' 2015.

\bibitem{li2014chart}
H.~Li and N.~Moacdieh, ``Is ``chart junk'' useful? an extended examination of
  visual embellishment,'' in {\em Proceedings of the Human Factors and
  Ergonomics Society Annual Meeting}, vol.~58, pp.~1516--1520, Sage
  Publications Sage CA: Los Angeles, CA, 2014.

\bibitem{bateman2010useful}
S.~Bateman, R.~L. Mandryk, C.~Gutwin, A.~Genest, D.~McDine, and C.~Brooks,
  ``Useful junk? the effects of visual embellishment on comprehension and
  memorability of charts,'' in {\em Proceedings of the SIGCHI conference on
  human factors in computing systems}, pp.~2573--2582, 2010.

\bibitem{correll2020truncating}
M.~Correll, E.~Bertini, and S.~Franconeri, ``Truncating the y-axis: Threat or
  menace?,'' in {\em Proceedings of the 2020 CHI Conference on Human Factors in
  Computing Systems}, pp.~1--12, 2020.

\bibitem{king2018preparing}
L.~King, ``Preparing better graphs,'' {\em // Journal of Public Health and
  Emergency}, vol.~2, no.~1, 2018.

\bibitem{witt2019graph}
J.~K. Witt, ``Graph construction: An empirical investigation on setting the
  range of the y-axis,'' {\em // Meta-psychology}, 2019.

\end{thebibliography}
\bibliographystyle{ieeetr}



\end{document}